\documentclass[journal]{IAENGtran}
\usepackage{textcomp}

\usepackage{xcolor}
\ifCLASSINFOpdf
   \usepackage[pdftex]{graphicx}
   \DeclareGraphicsExtensions{.pdf,.jpeg,.png}
\else
   \usepackage[dvips]{graphicx}
   \DeclareGraphicsExtensions{.eps}
\fi
\usepackage{rotating}
\usepackage{array}
\usepackage{dcolumn}
\usepackage{booktabs}
\usepackage{multirow}
\usepackage{bigstrut}
\usepackage{bigdelim}

\usepackage[utf8]{inputenc}
\usepackage[english]{babel}

\begin{document}
%\title{Social Cloud: Review and Roadmap from Models to Framework}
\title{A Critical Note on Social Cloud}

\author{Pramod~C.~Mane,~\IAENGmembership{}
        Kapil~Ahuja,\IAENGmembership{}
	and~Pradeep~Singh,~\IAENGmembership{}
        \thanks{ %(Write the date on which you submitted your paper for review.) 
%This work was supported in part by the U.S. Department of Commerce under Grant BS123456 (sponsor and financial support acknowledgment goes here). Paper titles should be written in uppercase and lowercase letters, not all uppercase. Avoid writing long formulas with subscripts in the title; short formulas that identify the elements are fine.
}
\thanks{Pramod C. Mane and Pradeep Singh are with the Department of Computer Science and Engineering, National Institute of Technology, Raipur, 492010 India, e-mail: \{pmane, psingh\}.cs@nitrr.ac.in.}% <-this % stops a space
\thanks{Kapil Ahuja is with the Department of Computer Science and Engineering, Indian Institute of Technology, Indore, 453552 India, e-mail: kahuja@iiti.ac.in}}% <-this % stops a space

\maketitle

\pagestyle{empty}
\thispagestyle{empty}

\begin{abstract}
%\boldmath

The idea of a social cloud has emerged as a resource sharing paradigm in a social network context. Undoubtedly, state-of-the-art social cloud systems demonstrate the potential of the social cloud acting as complementary to other computing paradigms such as the cloud, grid, peer-to-peer and volunteer computing. However, in this note, we have done a critical survey of the social cloud literature and come to the conclusion that these initial efforts fail to offer a general framework of the social cloud, also, to show the uniqueness of the social cloud. This short note reveals that there are significant differences regarding the concept of social cloud, resource definition, resource sharing and allocation mechanism, and its application and stakeholders. This study is an attempt to express a need for a general framework of the social cloud, which can incorporate various views and resource sharing setups discussed in the literature.

\end{abstract}
% IAENGtran.cls defaults to using nonbold math in the Abstract.
% of the abstract to achieve this. Many IAENG journals frown on math

\begin{IAENGkeywords}
social-cloud, cloud computing, grid computing, peer-to-peer computing, volunteer computing, network-services.
\end{IAENGkeywords}

% \fi
%
% For peerreview papers, this IAENGtran command inserts a page break and
\IAENGpeerreviewmaketitle

\section{Introduction}

\IAENGPARstart{I}{n} the past few years, researchers have shown an increased interest in real-world social relationships to develop the theory of social cloud. Till date, a number of research articles \cite{transaction, conference, socialcloudsconcept, transitioneconomy, colleresearch, publicresearch, distributed, healthsocialcloud, socialcloud-as-communitycloud, joballocationscoialcloud} have reported various socially based resource sharing setups under the tag of social cloud. 
The above numerous proof of the concepts bring forward social cloud concept through different architectural prototypes, deployment methods and shareable entities, principle stakeholders and so on. Indeed, these proposed concepts are discrete and dissimilar in nature. The literature reveals that these various resource sharing setups labelled as the social cloud have further appeared in various computing paradigms such as community cloud \cite{nist-new, nistcloud}, grid computing \cite{anatomy}, peer-to-peer (P2P), volunteer computing \cite{volunteer} and network services \cite{boyd-SNS}. 

Although majority of researchers have demonstrated the potential of the social cloud by keeping the social relationship as a central theme, the discrete and dissimilar views do not provide a unified view of the social cloud. Consequently, it resulted in confusion about the concept of social cloud. 

Our goal is to provide a critical survey of social cloud theory to show how the concept of social cloud is variegated. For this, we identify some points from the theory of social cloud that have contributed to distributed computing. We juxtapose social cloud with cloud, grid, P2P, and volunteer computing. The topics that we consider are architectures, resource definition, resource provision mechanisms, domains of applicability and associated principal stakeholders. We conclude with a general assessment concerning the need for a framework for social cloud. We believe that this type of effort will provide insight towards enhancing our knowledge about the social cloud, and further, will help us to find the research gap in this field as well help us taking necessary action to fill the gap.

%----------------------------------------
\begin{table*}[h!]
\footnotesize
\begin{center}
\caption{Different social cloud views *\footnotesize{Here paradigm means Social Cloud}, **\footnotesize{Healthcare Social Cloud}
}
%\begin{sideways}
\begin{tabular}{ p{2.9cm} p{11.6cm}  }\toprule
\textbf{Authors} & \textbf{View Social Cloud as} \\ \bottomrule
Chard et al. \cite{transaction} & `...\textit{Social Cloud is not representative of point-to-point ex-changes between users, rather it represents multipoint sharing within a whole community group}...' \\ \hline
John et al. \cite{publicresearch} & `...\textit{One way of thinking about the Social Cloud is to consider that social network groups are analogous to dynamic Virtual Organizations (VOs) \cite{anatomy}}...'. \\ \hline
Thaufeeg et al. \cite{colleresearch} & \textit{``Social Cloud Computing is a resource sharing framework in which resources and services are shared amongst individuals on the premise of the relationships and policies encoded in a social network.''} \\ \hline
Mohaisen et al. \cite{distributed} & `...\textit{Our paradigm* and model are similar in many aspects to the conventional grid-computing paradigm}...'.\\ \hline
Wooten et al. \cite{healthsocialcloud} & `...\textit{we describe our design** and prototype implementation of a social healthcare network over the cloud}...'. \\ \hline
Caton et al. \cite{socialcloud-as-communitycloud} & \textit{``A social cloud is a form of community cloud (as defined in NIST\textquotesingle s definition of cloud computing \cite{nistcloud}), as the resources are owned, provided and consumed by members of a social community''}. \\ \hline
Zhang et al. \cite{joballocationscoialcloud} & `...\textit{social cloud systems are constructed with peer-to-peer architectures with resources being owned and managed distributedly by individual users}...' \\ \hline
Xu et al. \cite{socialcloudsconcept} &  `...\textit{Social Clouds are owned and operated by the contributors or providers, who are also Social Clouds users. This is in contrast to Public Clouds where the platforms underlying a cloud are often owned and managed by a single service provider, and to Private Clouds where the platforms underlying a cloud are often owned and managed by a single enterprise}...' \\ \hline
Rafael Pezzi \cite{transitioneconomy} & \textit{``You may think of the Social Cloud concept as a blend of an auction web-site and a social networking website hosted at a cloud computing environment, which on its turn is running in a peer-to-peer network''} \\ \bottomrule
\end{tabular}
%\end{sideways}

\label{table:social-cloud-views}
\end{center}
\end{table*}
%----------------------------------------

%%%%%%%%%%%%%%%%%%%%%%%%%%%%%%
\section{social cloud: State of The Art}\label{current-status}
%%%%%%%%%%%%%%%%%%%%%%%%%%%%%%
In this section, we overview the concept of social cloud by summarizing its definitions and systems discussed in the literature.
\begin{table*}[ht!]
\small
\centering
\caption{Summary of Some Existing social cloud Systems}
%\begin{sideways}
%\begin{tabular}{l p{8.7cm}}
\begin{tabular}{p{3.5cm} p{1.6cm} p{2.5cm} p{2cm} p{2.5cm} p{2.5cm}}
  \toprule[1.5pt]
\textbf{Social Cloud Model} & \textbf{Resource Definition} & \textbf{Stakeholder} & \textbf{Deployment} & \textbf{Application} & \textbf{Resource Trading Mechanism}\\  \hline
Social Storage Cloud \cite{transaction}   &  Storage  & Social Cloud Members  & Facebook Application  & Storage-as-Service &  Reverse Auction, Posted Price\\ \hline
CRB-Model \cite{snsharing}  &  Amazon EC2 & Social Cloud Members  & Facebook Application  & -  & Bartering
\\ \hline
Social Collaborative Cloud \cite{colleresearch,escience} & VM or Amazon S3  & Scientist  & Facebook Application  & Platform-as-Service (PaaS) & - \\ \hline
Distributed Computing \cite{distributed}  &  Storage  & Social Cloud Members  & Overlay on Social Graph  & Task Outsourcing & Altruistic \\ \hline 
Social Cloud for Public eResearch \cite{publicresearch, escience}  &  VM, Storage, CPU, etc.  & BOINC Project and Facebook User  & Amalgamation of Facebook \& BOINC  & Volunteer Computing on Social Network & - \\ \hline
Husky Healthcare \cite{healthsocialcloud}& Blogs &  Patients, Physicians & Networking Services & Health-Care &-\\
\bottomrule[1.3pt]
\end{tabular}
%\end{sideways}
\label{table:classification2}
\end{table*}
%-----------------------------
%**********************************************************************
\subsection{Social Cloud Concept}
%**********************************************************************
Although the study \cite{Chard-Retrospective} has elaborated the concept of social cloud systematically, it has missed out various views on the social cloud discussed in the literature. Table \ref{table:social-cloud-views} lists the different views (definitions) of social cloud. These views demonstrate that social cloud acts as cloud, grid and peer-to-peer, volunteer computing, and social networking services\footnote{We draw such outward forms of social cloud on how it is described while proposing a particular system and researchers' explicit argument.}.
Table \ref{table:social-cloud-views} shows that there is no general agreement on the concept of social cloud. Next, we describe widely discussed social cloud systems.

%---------------------------------------------
\subsection{Social Cloud Systems}
%---------------------------------------------
\subsubsection{Social Storage Cloud (SSC)}
%Chard et al. \cite{transaction, conference} presents SSC as the proof of the concept. 
SSC \cite{transaction, conference} is deployed as a Facebook application. SSC users offer storage as a service to their friends. SSC applies posted price and/ or reverse auction market economics models for resource allocation. The Facebook application serves as a marketplace (so that users can communicate with each other and hence resource trading can be possible), currency regulation, and service level agreement between a provider and a consumer.
\subsubsection{Social Cloud for Public eResearch (SCPeR)}\label{sub:publicresearch}
SCPeR \cite{publicresearch} is a public e-research system that integrates BOINC \cite{boinc} platform\footnote{http://boinc.berkeley.edu/} and Facebook so that social network users are able to donate computational resources (e.g., processing power, storage, etc.) to various scientific projects (for example SETI@Home\footnote{http://setiathome.ssl.berkeley.edu/}, etc).  
The SCPeR users and scientific projects view SCPeR system differently. From the point of view of users, SCPeR is a Facebook application, whereas form perspective of projects servers, it is an account management system. 
\subsubsection{Social Collaborative Cloud}
SoCC \cite{colleresearch} aims to enable computational resource sharing within a scientific community. Here, the scientific community refers to a group of either researchers or institutions. The architecture of SoCC allows researchers or institutions to share their sparse computing resources (in the form of Virtual Machines) on rent. 
\subsubsection{Social Cloud (SC)}
Mohaisen et al. \cite{distributed} presents SC as a distributed computing paradigm where social network users collaboratively construct a pool of storage resources. SC users share their computing resource sharing altruistically with only their immediate neighbours. In SC, a user can outsource a computing task to its neighbours, and then, the neighbours perform the outsourced task altruistically on behalf of the task owner. 
\subsubsection{Husky Healthcare Social Cloud (HHSC)} 
HHSC \cite{healthsocialcloud} is a cloud-assisted social health care network. It facilitates standard users (either patients or the users who want to get health knowledge) and health professionals to share health-related information with each other through blogs. In HHSC, standard users can create, update and delete posts and comments to their blogs. A standard user can read and comments on all blogs owned by other standard users. Health professionals read posts and comments of all blogs and create comments on all posts.
\subsubsection{Cloud Resource Bartering (CRB) Model}
CRB model of social cloud \cite{snsharing} enables social network users for sharing a part of their owned Amazon EC2 infrastructure with other users. Users perform such resource sharing in the network through the bartering model.
%-----------------------------------------
\subsubsection{Other social cloud settings}
%-----------------------------------------
Zhang et al. \cite{joballocationscoialcloud} present Cloud-Operator-Based Social Cloud (CObSC) system where a user can outsource a task and requested payment to the cloud operator. The cloud operator is the owner of the market. A buyer submits its job to the operator. Then, the operator do the following: First, the operator breaks the submitted job into small tasks. Second, it selects a set of capable users (who are part of the system and act as suppliers at the given time). Third, collection of tasks are assigned to the selected suppliers to perform the task. After completion of the assigned task by an individual supplier, the operator makes payment to the suppliers. Ref. \cite{socialcloudsconcept} presents the concept of social cloud as an alternative to public and private clouds. In this case, participants collaborate to build a cloud platform for achieving their goals. Further, such a cloud platform neither has single ownership nor is controlled by a single entity. The above setup is contrary to public and private clouds (as these cloud platforms are managed by a single service provider or by a single enterprise, respectively).

%**********************************************************************
\section{Discussion}
%**********************************************************************
In this section, we critically overview the concept of the social cloud. For this, we consider the aspects such as views of social cloud, shareable entity, the scope of resource sharing, resource sharing/ trading mechanism, and applications. As social cloud is appeared in the form of a cloud, grid, P2P, volunteer computing and network services, we make use of comparison based analysis of the above-listed computing frameworks for critical examination of social cloud concept.

%###############################
\subsection{Social Cloud: Different Views}
%################################
What social cloud exactly is? This is the fundamental question that arises from the existing literature. As mentioned earlier that there is no agreement on what exactly social cloud is. Researchers have projected the idea of social cloud in various ways (discussed in the previous section; see Table \ref{table:classification2}), which eventually leads to not only confusion but also raises important questions regarding the  uniqueness of the social cloud. %from the other computing frameworks (listed above). 
Firstly, how can the social cloud be in the form of the cloud, grid, peer-to-peer and volunteer computing, and network services at the same time? Given that, these computing frameworks, although share a few common characteristics, they can be distinguishable with respect to their architectures and several other parameters\footnote{For a more detailed comparison-based surveys on cloud, grid, peer-to-peer, and volunteer computing, and network services, we refer the reader to \cite{foster-p2pvsgrid, cabani-p2pvsgrid, cloudgrid360, kondo-gridvscloud, zhang-gridvscloud, brandic-gridvscloud, sadashiv-gridvscloud}.}. For instance, grid and peer-to-peer computing can be compared on the basis of target communities, incentives, resources, applications and scalability \cite{foster-p2pvsgrid, cabani-p2pvsgrid}. Grid and cloud computing can be compare on the basis of business model, architecture, resource management, programming model, application model, security model, and cost-benefit analysis \cite{cloudgrid360, kondo-gridvscloud}. Another concern is that can we consider social cloud as a unique computing framework like the other computing frameworks. If we exclude those views which suppose social cloud as a networking service and volunteer computing, can we simply argue that social cloud is a case of jungle computing \cite{junglecomputing} (which comprises cloud, grid, and, cluster computing) in a social network context?

%***************************************************
\subsection{Social Cloud: Resources and Trading Mechanism}
%***************************************************
One can observe that resource sharing is a vital factor in the social cloud. But the literature fails to answer the concerns like what resources are to be considered as a shareable entity and what resource sharing/ trading protocols are available? 

In the social cloud, resources are multitudinous and could be information, computing resources (e.g., storage, computing power, etc.), people, software licenses, blog, workflow or the part of the cloud (e.g., Amazon EC2) (see Table \ref{table:classification2}). On the contrary, cloud, grid, P2P, volunteer computing frameworks are specific about resource definition and their properties. For example, in a grid system unified more powerful computing resources include cluster, storage resources, scientific instruments or data sets. P2P and volunteer computing integrate personal computers available at the Internet edge. Cloud is a different computing paradigm that offers \textbf{S}-as-service (\textbf{S} stands for infrastructure, data, software, platform, information, or network). %Unlike, these computing frameworks, the resource definition in the social cloud is multitudinous (see Table \ref{table:classification2}). 

In the social cloud, if {\it people} are considered to be a resource \cite{transaction} then how is the social cloud is different from crowdsourcing. Next, there are several online social networking sites providing a platform for users to share textual, audio, video information with others, and hence, can we view these online social networking sites as a social cloud? If so, can we differentiate social cloud and networking services? The above questions impose us to define resources and their properties in the social cloud context. 

Unlike, cloud, grid or volunteer computing, the social cloud does not follow a single standard resource sharing/trading or allocation mechanism. For instance, cloud computing follows {\it pay as you go model} for resource allocation. In grid computing, VO members first build a pool of resources, and then allocate those resources to the one who need it through a predefined protocol set by the members.

%###############################
\subsection{Social Cloud: Scope}
%################################
An another concern is that can we define the scope of social cloud in terms of resource sharing?
%Research and development on Social Cloud is currently at its infancy. In the literature, there seems to be no general definition of Social Cloud (see Table \ref{table:social-cloud-views}). 
The different views, as seen in Table \ref{table:social-cloud-views}, indicates that nearly half of the studies \cite{ transaction, conference, colleresearch, publicresearch, distributed, socialcloud-as-communitycloud} consider the social cloud as a social network group analogous to the dynamic Virtual Organization (VO) \cite{anatomy} in a social network context. A VO like social network group is a set of users who collaboratively pool resources to achieve its certain common goal. A group can achieve its common goal by defining a set of policies regarding group membership, resource sharing, and so on. So one can say that social cloud paradigm stands on the notion of collaboration between generally smaller, better-connected groups of social network users with more heterogeneous resources to share. The above kind of resource sharing depicts {\it local sharing}.
\par On the other side, a few studies \cite{socialcloudsconcept, transitioneconomy, healthsocialcloud, joballocationscoialcloud} view the social cloud as a form of peer-to-peer community (or peer-to-peer social networking) where each participant performs the same role as the others and each participant manages its own resources at its end. A participant who joins the network provides resources to others and avails resources provided by others. In this way, the social cloud is neither subject to centralized control nor owned by a single entity. So one can view social cloud paradigm as a collaboration between a generally bigger, loosely connected group of users present at the edge of Internet sharing varied types of resources. The above kind of resource sharing gives an expression of {\it global sharing}.
%***************************************************
\subsection{Social Cloud: Target Communities and Stakeholders}
%***************************************************
Next we look at the targeted communities and stakeholders of the social cloud? Unlike the listed computing frameworks, in the social cloud, the participants and beneficiaries are not well understood. For example, in cloud computing, the participants and beneficiaries are defined by its deployment methods like public, private, hybrid, or community cloud. In grid computing, it is a virtual organization, whereas, in volunteer computing, the participants are users at Internet-edge and beneficiaries are scientific projects. According to the widely accepted definition of the social cloud\footnote{Chard et al. \cite{transaction} define \textit{``A social cloud is a resource and service sharing framework utilizing relationships established between members of a social network''.}}, the participants and beneficiaries are social network members (socially connected) who form a group like VO. However, many social cloud applications comprise diverse and anonymous participants and beneficiaries. For instance, in SCPeR \cite{publicresearch}, the participants are social network users (socially connected), whereas, the beneficiaries are scientific projects. In some cases, the systems do not utilise  relationships established between the members of a social network \cite{healthsocialcloud, bidoki-bee-colony-algorithm, networkservices}. This contradicts the social cloud definition in \cite{transaction}.

\section{Conclusion}
Distributed computing frameworks like cloud, grid, P2P, and volunteer are well defined in terms of architecture, deployment methods, resources and principle stakeholders. On the contrary, early-stage research efforts in the social cloud do not provide the actual system, and hence, fall short of defining it as the above-listed computing paradigms. In a nutshell, the existing literature fails to provide; a clear vision of the social cloud, an architecture (that incorporates all the views discussed in the literature), its resource and trading mechanism, a clear understanding of its scope, its target communities and beneficiaries, and applicability.

A strong debate going on is whether the social cloud is an alternative to other distributed computing such as the cloud, grid, P2P and volunteer computing. We are of the opinion that these are not competing paradigm, rather, they have their own merits as well as limitations. %More precisely, the social cloud should be dealt with different research care and development so that ultimately it can emerge as another fruitful distributed computing paradigm. 
More precisely, the social cloud should be dealt with different research focuses. Issues like security, reliability and scalability of user resources deployed in the social cloud infrastructure are the major concerns that need immediate research attention.

We admit this study is not exhaustive rather it provides a glimpse of the current state of the art by summarizing existing research efforts and a critical review. We believe that the article will help the readers to examine the space between what is the existing status of the field and what needs to be addressed. Finally, we argue that there is a need for a general framework that would incorporate all the views on social cloud discussed in the literature, and further, offers the clear concept of the social cloud, its scope and limitations, resources and resource sharing/trading mechanisms, and its applicability and targeted communities.


\begin{thebibliography}{10}
\providecommand{\url}[1]{#1}
\csname url@samestyle\endcsname
\providecommand{\newblock}{\relax}
\providecommand{\bibinfo}[2]{#2}
\providecommand{\BIBentrySTDinterwordspacing}{\spaceskip=0pt\relax}
\providecommand{\BIBentryALTinterwordstretchfactor}{4}
\providecommand{\BIBentryALTinterwordspacing}{\spaceskip=\fontdimen2\font plus
\BIBentryALTinterwordstretchfactor\fontdimen3\font minus
  \fontdimen4\font\relax}
\providecommand{\BIBforeignlanguage}[2]{{%
\expandafter\ifx\csname l@#1\endcsname\relax
\typeout{** WARNING: IEEEtran.bst: No hyphenation pattern has been}%
\typeout{** loaded for the language `#1'. Using the pattern for}%
\typeout{** the default language instead.}%
\else
\language=\csname l@#1\endcsname
\fi
#2}}
\providecommand{\BIBdecl}{\relax}
\BIBdecl

\bibitem{transaction}
K.~Chard, K.~Bubendorfer, S.~Caton, and O.~Rana, ``Social cloud computing: A
  vision for socially motivated resource sharing,'' \emph{IEEE Transactions on
  Services Computing}, vol.~5, no.~4, pp. 551--563, Fourth 2012.

\bibitem{conference}
K.~Chard, S.~Caton, O.~Rana, and K.~Bubendorfer, ``Social cloud: Cloud
  computing in social networks,'' in \emph{3rd IEEE International Conference on
  Cloud Computing (CLOUD)}, July 2010, pp. 99--106.

\bibitem{socialcloudsconcept}
S.~Xu and M.~Yung, ``Social Clouds: Concept, security architecture and some
  mechanisms,'' in \emph{Proceedings of the First International Conference on
  Trusted Systems}, ser. INTRUST'09.\hskip 1em plus 0.5em minus 0.4em\relax
  Springer-Verlag, 2010, pp. 104--128.

\bibitem{transitioneconomy}
\BIBentryALTinterwordspacing
R.~Pezzi, ``Information technology tools for a transition economy,'' September
  2009. [Online]. Available:
  \url{http://www.socialcloud.net/papers/ITtools.pdf}
\BIBentrySTDinterwordspacing

\bibitem{colleresearch}
A.~Thaufeeg, K.~Bubendorfer, and K.~Chard, ``Collaborative e{R}esearch in a
  social cloud,'' in \emph{7th IEEE International Conference on E-Science
  (e-Science)}, Dec 2011, pp. 224--231.

\bibitem{publicresearch}
K.~John, K.~Bubendorfer, and K.~Chard, ``A social cloud for public
  e{R}esearch,'' in \emph{7th IEEE International Conference on E-Science
  (e-Science).}, Dec 2011, pp. 363--370.

\bibitem{distributed}
A.~Mohaisen, H.~Tran, A.~Chandra, and Y.~Kim, ``Trustworthy distributed
  computing on social networks,'' in \emph{Proceedings of the 8th ACM SIGSAC
  Symposium on Information, Computer and Communications Security}, ser. ASIA
  CCS '13.\hskip 1em plus 0.5em minus 0.4em\relax ACM, 2013, pp. 155--160.

\bibitem{healthsocialcloud}
R.~Wooten, R.~Klink, F.~Sinek, Y.~Bai, and M.~Sharma, ``Design and
  implementation of a secure healthcare social cloud system,'' in \emph{12th
  IEEE/ACM International Symposium on Cluster, Cloud and Grid Computing
  (CCGrid)}, May 2012, pp. 805--810.

\bibitem{socialcloud-as-communitycloud}
S.~Caton, C.~Haas, K.~Chard, K.~Bubendorfer, and O.~Rana, ``A social compute
  cloud: Allocating and sharing infrastructure resources via social networks,''
  \emph{IEEE Transactions on Services Computing}, vol.~7, no.~3, pp. 359--372,
  July 2014.

\bibitem{joballocationscoialcloud}
Y.~Zhang and M.~van~der Schaar, ``Incentive provision and job allocation in
  social cloud systems,'' \emph{IEEE Journal on Selected Areas in
  Communications}, vol.~31, no.~9, pp. 607--617, September 2013.

\bibitem{nist-new}
L.~Badger, T.~Grance, R.~Patt-Corner, and J.~Voas, ``Draft cloud computing
  synopsis and recommendations,'' \emph{NIST special publication}, vol. 800, p.
  146, 2011.

\bibitem{nistcloud}
M.~Hogan, F.~Liu, A.~Sokol, and J.~Tong, ``NIST cloud computing standards
  roadmap,'' National Institute of Standards and Technology, Tech. Rep., 2011.

\bibitem{anatomy}
I.~Foster, C.~Kesselman, and S.~Tuecke, ``The anatomy of the grid: Enabling
  scalable virtual organizations,'' \emph{Int. J. High Perform. Comput. Appl.},
  vol.~15, no.~3, pp. 200--222, Aug. 2001.

\bibitem{volunteer}
L.~F. Sarmenta, ``Volunteer computing,'' Ph.D. dissertation, Massachusetts
  Institute of Technology, 2001.

\bibitem{boyd-SNS}
d.~m. boyd and N.~B. Ellison, ``{Social Network Sites: Definition, History, and
  Scholarship},'' \emph{Journal of Computer-Mediated Communication}, vol.~13,
  no.~1, pp. 210--230, 10 2007.

\bibitem{snsharing}
Z.~Ali, R.~Rasool, and P.~Bloodsworth, ``Social networking for sharing cloud
  resources,'' in \emph{Second International Conference on Cloud and Green
  Computing (CGC)}, Nov 2012, pp. 160--166.

\bibitem{escience}
K.~Bubendorfer, K.~Chard, K.~John, and A.~M. Thaufeeg, ``e{S}cience in the
  social cloud,'' \emph{Future Generation Computer Systems}, vol.~29, no.~8,
  pp. 2143 -- 2156, 2013.

\bibitem{Chard-Retrospective}
K.~{Chard}, S.~{Caton}, O.~{Rana}, and K.~{Bubendorfer}, ``Social clouds: A
  retrospective,'' \emph{IEEE Cloud Computing}, vol.~2, no.~6, pp. 30--40,
  2015.

\bibitem{boinc}
D.~P. Anderson, ``Boinc: A system for public-resource computing and storage,''
  in \emph{Proceedings of the 5th IEEE/ACM International Workshop on Grid
  Computing}, ser. GRID '04.\hskip 1em plus 0.5em minus 0.4em\relax IEEE
  Computer Society, 2004, pp. 4--10.

\bibitem{foster-p2pvsgrid}
I.~Foster and A.~Iamnitchi, ``On death, taxes, and the convergence of
  peer-to-peer and grid computing,'' in \emph{Peer-to-Peer Systems II},
  M.~Kaashoek and I.~Stoica, Eds.\hskip 1em plus 0.5em minus 0.4em\relax
  Springer Berlin Heidelberg, 2003, vol. 2735, pp. 118--128.

\bibitem{cabani-p2pvsgrid}
A.~Cabani, S.~Ramaswamy, M.~Itmi, S.~Al-Shukri, and J.~Pécuchet, ``Distributed
  computing systems: P2P versus grid computing alternatives,'' in
  \emph{Innovations and Advanced Techniques in Computer and Information
  Sciences and Engineering}, T.~Sobh, Ed.\hskip 1em plus 0.5em minus
  0.4em\relax Springer Netherlands, 2007, pp. 47--52.

\bibitem{cloudgrid360}
I.~Foster, Y.~Zhao, I.~Raicu, and S.~Lu, ``Cloud computing and grid computing
  360-degree compared,'' in \emph{Grid Computing Environments Workshop, 2008.
  GCE '08}, Nov 2008, pp. 1--10.

\bibitem{kondo-gridvscloud}
D.~Kondo, B.~Javadi, P.~Malecot, F.~Cappello, and D.~Anderson, ``Cost-benefit
  analysis of cloud computing versus desktop grids,'' in \emph{IEEE
  International Symposium on Parallel Distributed Processing,(IPDPS)}, May
  2009, pp. 1--12.

\bibitem{zhang-gridvscloud}
S.~Zhang, X.~Chen, S.~Zhang, and X.~Huo, ``The comparison between cloud
  computing and grid computing,'' in \emph{International Conference on Computer
  Application and System Modeling (ICCASM)}, vol.~11, Oct 2010, pp.
  V11--72--V11--75.

\bibitem{brandic-gridvscloud}
I.~Brandic and S.~Dustdar, ``Grid vs cloud-a technology comparison,''
  \emph{it-Information Technology Methoden und innovative Anwendungen der
  Informatik und Informationstechnik}, vol.~53, no.~4, pp. 173--179, 2011.

\bibitem{sadashiv-gridvscloud}
N.~Sadashiv and S.~Kumar, ``Cluster, grid and cloud computing: A detailed
  comparison,'' in \emph{6th International Conference on Computer Science
  Education (ICCSE)}.\hskip 1em plus 0.5em minus 0.4em\relax IEEE, Aug 2011,
  pp. 477--482.

\bibitem{junglecomputing}
F.~Seinstra, J.~Maassen, R.~van Nieuwpoort, N.~Drost, T.~van Kessel, B.~van
  Werkhoven, J.~Urbani, C.~Jacobs, T.~Kielmann, and H.~Bal, ``Jungle computing:
  Distributed supercomputing beyond clusters, grids, and clouds,'' in
  \emph{Grids, Clouds and Virtualization}, M.~Cafaro and G.~Aloisio, Eds.\hskip
  1em plus 0.5em minus 0.4em\relax Springer London, 2011, pp. 167--197.

\bibitem{bidoki-bee-colony-algorithm}
M.~Z. Bidoki and M.~J. Kargar, ``A social cloud computing: Employing a bee
  colony algorithm for sharing and allocating tourism resources,'' \emph{Modern
  Applied Science}, vol.~10, no.~5, p. 177, 2016.

\bibitem{networkservices}
M.~Sato, ``Creating next generation cloud computing based network services and
  the contributions of social cloud operation support system (oss) to
  society,'' in \emph{18th IEEE International Workshops on Enabling
  Technologies: Infrastructures for Collaborative Enterprises, (WETICE '09)},
  June 2009, pp. 52--56.

\end{thebibliography}
\end{document}